\documentclass[12pt,preprint]{aastex}
\usepackage{hyperref}
\usepackage{natbib}
\usepackage{textcomp,gensymb}

\newcommand{\sdo}{{\it SDO}}
\newcommand{\goes}{{\it GOES}}
\newcommand{\rhessi}{{\it RHESSI}}

\begin{document}
	
\title{Direct Observation of Two-Step Magnetic Reconnection in a Solar Flare}

\author{Tingyu Gou\altaffilmark{1,2}, Astrid M. Veronig\altaffilmark{2}, Ewan Dickson\altaffilmark{2}, Aaron Hernandez-Perez\altaffilmark{2}, Rui Liu\altaffilmark{1,3}}

\altaffiltext{1}{CAS Key Laboratory of Geospace Environment, Department of Geophysics and Planetary Sciences, University of Science and Technology of China, Hefei 230026, China; tygou@mail.ustc.edu.cn}
\altaffiltext{2}{Kanzelh\"{o}he Observatory \& Institute of Physics/IGAM, University of Graz, Universit\"{a}tsplatz 5, 8010 Graz, Austria}
\altaffiltext{3}{Collaborative Innovation Center of Astronautical Science and Technology, Hefei 230026, China}
	
\begin{abstract}

We report observations of an eruptive X2.8 flare on 2013 May 13, which shows two distinct episodes of energy release in the impulsive phase. The first episode is characterized by the eruption of a magnetic flux rope, similar to the energy-release process in most standard eruptive flares. While the second episode, which is stronger than the first normal one and shows enhanced high-energy X-ray and even $\gamma$-ray emissions, is closely associated with magnetic reconnection of a large-scale loop in the aftermath of the eruption. The reconnection inflow of the loop leg is observed in the Solar Dynamics Observatory (\sdo)/Atmospheric Imaging Assembly (AIA) 304~\r{A} passband and accelerates towards the reconnection region to a speed as high as $\sim$130~km/s. Simultaneously the corresponding outflow jets are observed in the AIA hot passbands with a speed of $\sim$740~km/s and mean temperature of $\sim$14~MK. \rhessi\ observations show a strong burst of hard X-ray (HXR) and $\gamma$-ray emissions with hard electron spectra of $\delta\approx3$, exhibiting a soft-hard-harder behavior. A distinct altitude decrease of the HXR loop-top source coincides with the inward swing of the loop leg observed in the AIA 304~\r{A} passband, which is suggested to be related to the coronal implosion. This fast inflow of magnetic flux contained in the loop leg greatly enhances the reconnection rate and results in very efficient particle acceleration in the second-step reconnection, which also helps to achieve a second higher temperature peak up to T $\approx$ 30~MK. 

\end{abstract}
	

\section{Introduction}

Magnetic reconnection is widely considered as the fundamental energy release process in solar flares, which are among the most magnificent phenomena on the Sun. In the classical CSHKP flare model \citep{Carmichael1964, Sturrock1966, Hirayama1974, Kopp1976}, a prominence or flux rope moves outwards and magnetic reconnection continuously occurs in the current sheet region underneath, leaving behind rising soft X-ray (SXR) flare loops and separating ribbons. Due to pressure imbalance more field lines are swept into the reconnection region, which are generally referred to as inflows \citep[e.g.,][]{Yokoyama2001,Takasao2012,Savage2012}. Bi-directional outflows of the reconnection jets are presumed to move both upward and downward from the reconnection site, and are sometimes observed in different forms such as plasmoids, supra-arcade downflows and newly-formed loops \citep{Innes1997,Wang2007,Liu2013,LiuW2013,Su2013}. However the search for direct observations of these features is still needed because of the tight relation to magnetic reconnection.
 
The energy released by magnetic reconnection is used for plasma heating and particle acceleration \citep[see the reviews by ][]{Fletcher2011,Holman2016,Benz2017}, for which the Reuven Ramaty High-Energy Solar Spectroscopic Imager \citep[\rhessi;][]{Lin2002} provides both imaging \citep{Hurford2002} and spectral analyses \citep{Smith2002}. The observed double coronal sources \citep[e.g.,][]{Sui2003,Veronig2006,LiuW2008,Su2013} are considered as strong evidence for magnetic reconnection, and to relate to the heating in both upper and lower reconnection outflow regions. The loop-top (LT) X-ray source is observed to move downward in the early phase of flares before its commonly upward motion, for which there are interpretations of different scenarios such as the relaxation of newly reconnected field lines, particle acceleration in a collapsing magnetic trap and coronal implosion \citep{Sui2003,Sui2004,Veronig2006,Ji2007,Liu2009}. The HXR spectra in solar flares typically show a hard non-thermal distribution around the peak, and in most cases they soften again towards the flare end, exhibiting a soft-hard-soft (SHS) behavior \citep{Grigis2004}. However, there are also some flares for which the spectra continue to harden after the peak, referred to as a soft-hard-harder (SHH) behavior, and these flares are shown to be preferably associated with solar energetic particle (SEP) events \citep{Grayson2009}.

In this letter we present observations of an X2.8 flare on 2013 May 13 showing two-step magnetic reconnection. Besides the first common step characterized by the eruption of a magnetic flux rope resembling most standard eruptive flares, the second anomalous one is manifested as strong reconnection of a large-scale loop in the aftermath of the eruption, which moves inward at a speed as high as $\sim$130~km/s. Simultaneously hot outflows and large energy release with emissions from high-energy particles up to the $\gamma$-ray range are recorded. This event has been studied by \citet{MartinezO2014} and \citet{SaintH2014}, who concentrated on the unusual loop-prominence system in the aftermath of the flare that was observed in white light by the Solar Dynamics Observatory \citep[\sdo;][]{Pesnell2012}/Helioseismic and Magnetic Imager\citep[HMI;][]{Schou2012}, indicative of very high coronal densities of the order of $10^{12}$~cm$^{-3}$. In the present paper, we concentrate on the plethora of magnetic reconnection signatures related to the second burst of strong energy release, using observations by \sdo/Atmospheric Imaging Assembly \citep[AIA;][]{Lemen2012} and \rhessi.

\section{Observation and Analysis}\label{sec:obs} 

The event under study takes place on 2013 May 13 in NOAA active region 11748 near the northeast solar limb, where it produces more than ten flares on that day including two X-class events. Here we concentrate on the X2.8 flare which starts at 15:48~UT and peaks at 16:05~UT. At the early phase a magnetic flux rope erupts out accompanied by a jet-like structure, and the eruption leads to a fast halo coronal mass ejection (CME), with a velocity of $\sim$1850~km/s according to the {\it SOHO}/LASCO CME catalog \footnote{\url{http://cdaw.gsfc.nasa.gov/CME_list/}}. The flare is followed by a long gradual phase lasting more than four hours, during which an elongated current sheet and a cusp-shaped structure can be clearly seen above the post-flare loops. 

\subsection{EUV Observation}\label{ssec:euv}

\begin{figure}[ht]
  \plotone{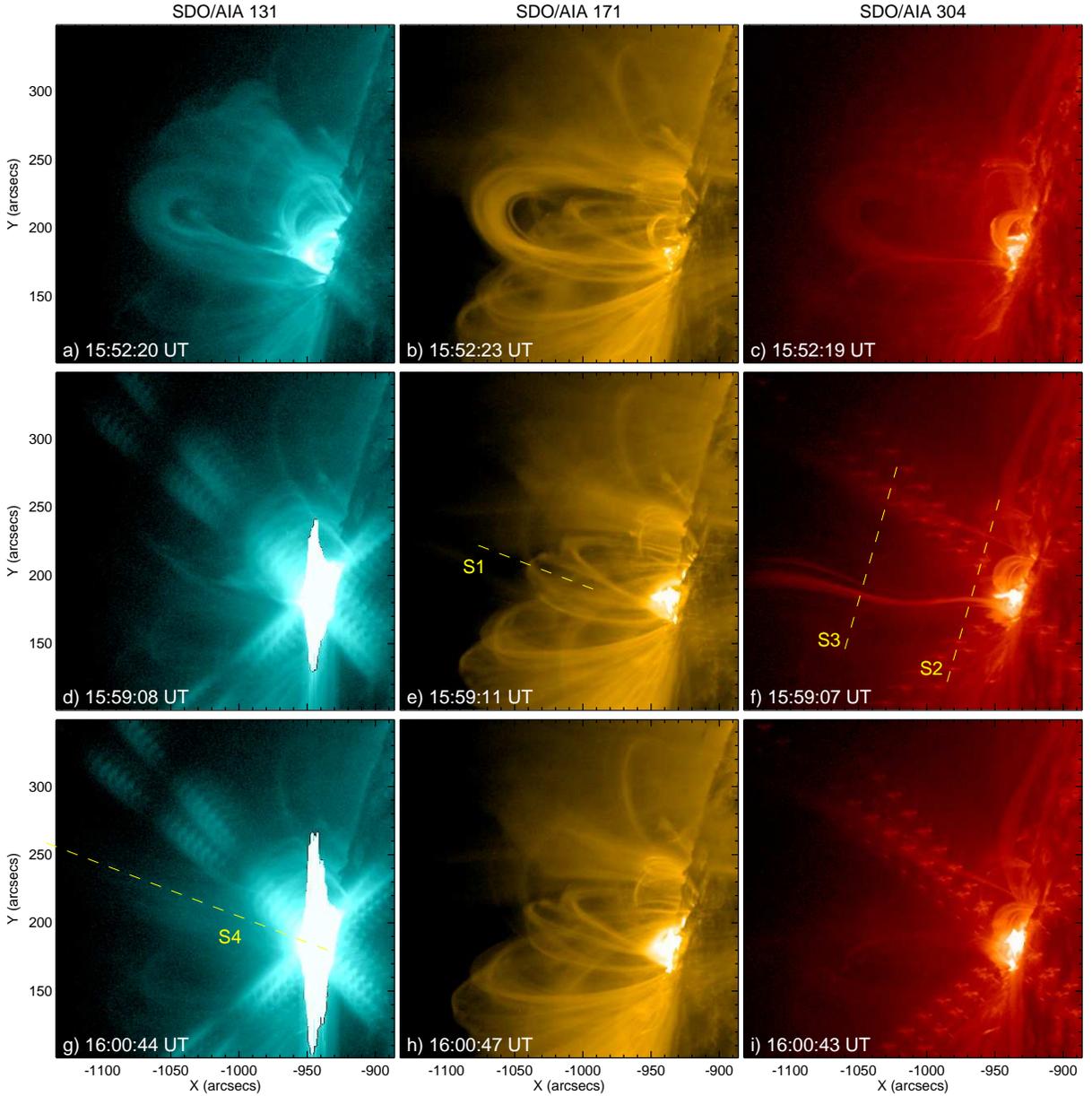}
  \caption{\small Snapshots of \sdo/AIA 131 (left), 171 (middle) and 304 (right)~\r{A} observations. The dashed lines in Panels (e--g) indicate the slits used to generate the stack plots in Figure \ref{fig:tempor} (d--g). An animation of this figure is available.
  \label{fig:maps}}
\end{figure}

\begin{figure}[ht]
  \centering
  \includegraphics[height=0.8\textheight]{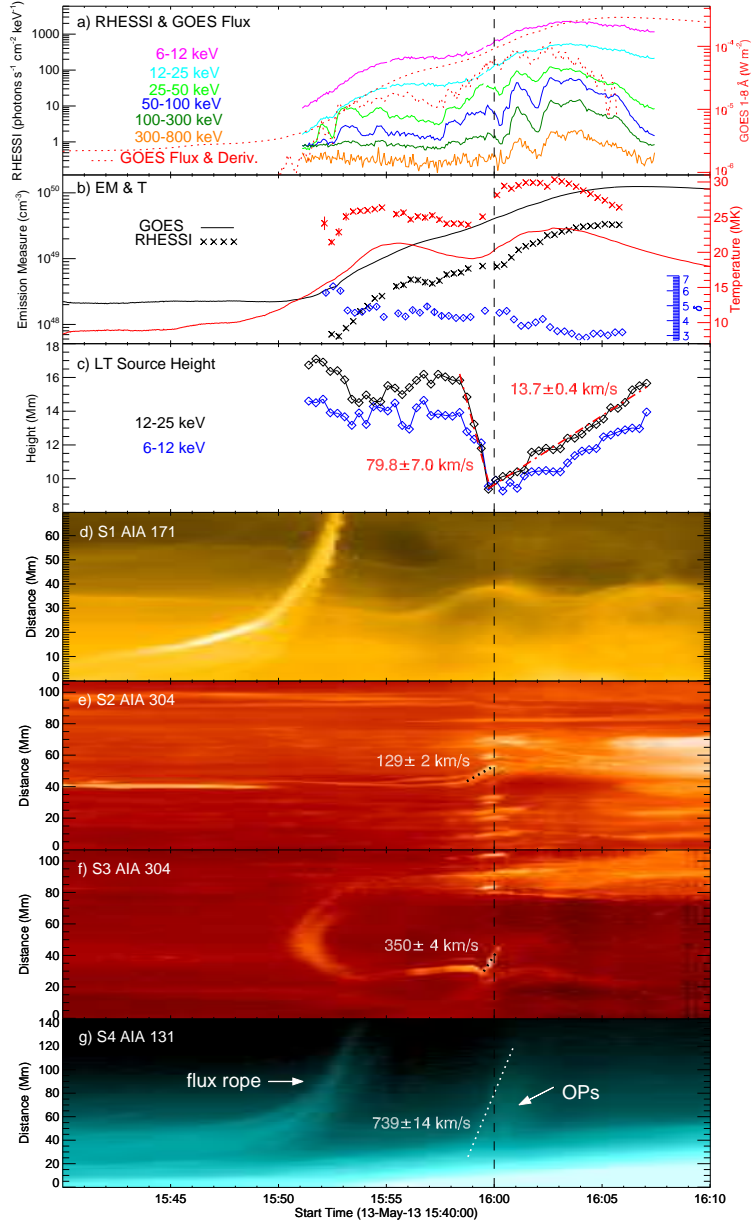}
  \caption{\small Temporal evolution of the impulsive phase. (a) \rhessi\ photon fluxes in different energy bands (scaled by the left y-axis, arbitrarily shifted vertically), \goes\ 1--8 \r{A} flux and its derivative (red dotted lines; scaled by the right y-axis, derivative is shifted arbitrarily). (b) Emission measure (black; scaled by the left y-axis) and temperature (red; scaled by the right y-axis) inferred from \rhessi\ (cross symbols with error bars) and \goes\ (solid lines) spectral fits, and the index of electron distribution derived from thick-target bremsstrahlung fits to the \rhessi\ spectra, $\delta$ (blue diamonds; scaled by an additional y-axis). (c) \rhessi\ LT source height evolution in 6--12 (blue) and 12--25 (black)~keV bands. Two linear fittings are plotted as red dashed lines. (d)--(g) Dynamic evolution seen through the slits in Figure \ref{fig:maps} (e--g) (see the text for details). The vertical dashed line marks the approximate time when the loop leg disappears, i.e., at 16:00~UT.
  \label{fig:tempor}}
\end{figure}

Figure \ref{fig:maps} shows the observations of \sdo/AIA three EUV passbands, i.e., 131~\r{A} (primarily contributed from Fe XXI line with a peak response temperature at $\log T = 7.05$), 171~\r{A} (Fe IX, $\log T = 5.85$) and 304~\r{A} (He II, $\log T = 4.7$). Some hot loops rise and expand rapidly in the early phase, which are likely associated with a magnetic flux rope. The erupting flux rope is connected to the flaring loop underneath by a linear feature (see AIA 131 \r{A} images in Figure \ref{fig:maps} and its animation), which is similar in location to the current sheet observed in the gradual phase. Some plasma blobs move upward along this structure at around 15:54~UT \citep[see the animation of Figure \ref{fig:maps}; for similar observations, see][]{Liu2010a,Liu2013,Zhu2016}. According to the standard flare model, this linear feature with highly dynamic characteristics most likely corresponds to the vertical current sheet that is formed in the wake of the erupting flux rope, where magnetic reconnection is supposed to occur. Some coronal loops are observed to contract toward the flaring region and oscillate after the eruption \citep[e.g.,][]{Liu2009,Liu2010,Simoes2013,Russell2015}, as seen in the AIA 171 \r{A} images in the animation of Figure \ref{fig:maps} and the stack plot in Figure \ref{fig:tempor}(d), which is generated by placing a virtual slit across the coronal loops (S1 in Figure \ref{fig:maps}(e), 155~pixels long and 6~pixels wide, measured from the low-altitude end).

The most intriguing phenomenon in this event is the behavior of a long leg-like structure observed in the AIA 304~\r{A} passband (see Figure \ref{fig:maps}(f)). It shows up in the cool EUV passbands as one leg of the stretched overlying loops, but becomes more and more prominent as the erupting flux rope moves outside from the AIA field of view (FOV). Then it is only detectable in the AIA 304~\r{A} passband, indicating a low temperature of $\sim$0.05~MK. At $\sim$15:59~UT the leg suddenly accelerates toward the region above the flaring loops and disappears (see the animation of Figure \ref{fig:maps}). To study the dynamics, we place two virtual slits across the loop leg (S2 and S3 in Figure \ref{fig:maps}(f), both are 240~pixels long and 6~pixels wide, measured from the south end). The time-distance plots in Figure \ref{fig:tempor}(e,f) show that after $\sim$15:59~UT the loop leg undergoes a fast swing northwards and disappears. The linear fits give the swing speeds of $\sim$130~km/s and $\sim$350~km/s respectively, with the upper part generally faster.

Almost at the same time when the loop leg disappears, some diffusive plasmas quickly move upward, which are only visible in the AIA hot passbands (e.g., 131~\r{A} image in Figure \ref{fig:maps}(g)). The virtual slit S4 of 325~pixels long and 10~pixels wide is placed on the 131~\r{A} images along these outward-moving plasmas (OPs) and measured from the low-altitude end. The stack plot in Figure \ref{fig:tempor}(g) shows the erupting flux rope as well as the diffusive OPs, with the out-moving speed of $\sim$740~km/s.

\begin{figure}[ht]
  \plotone{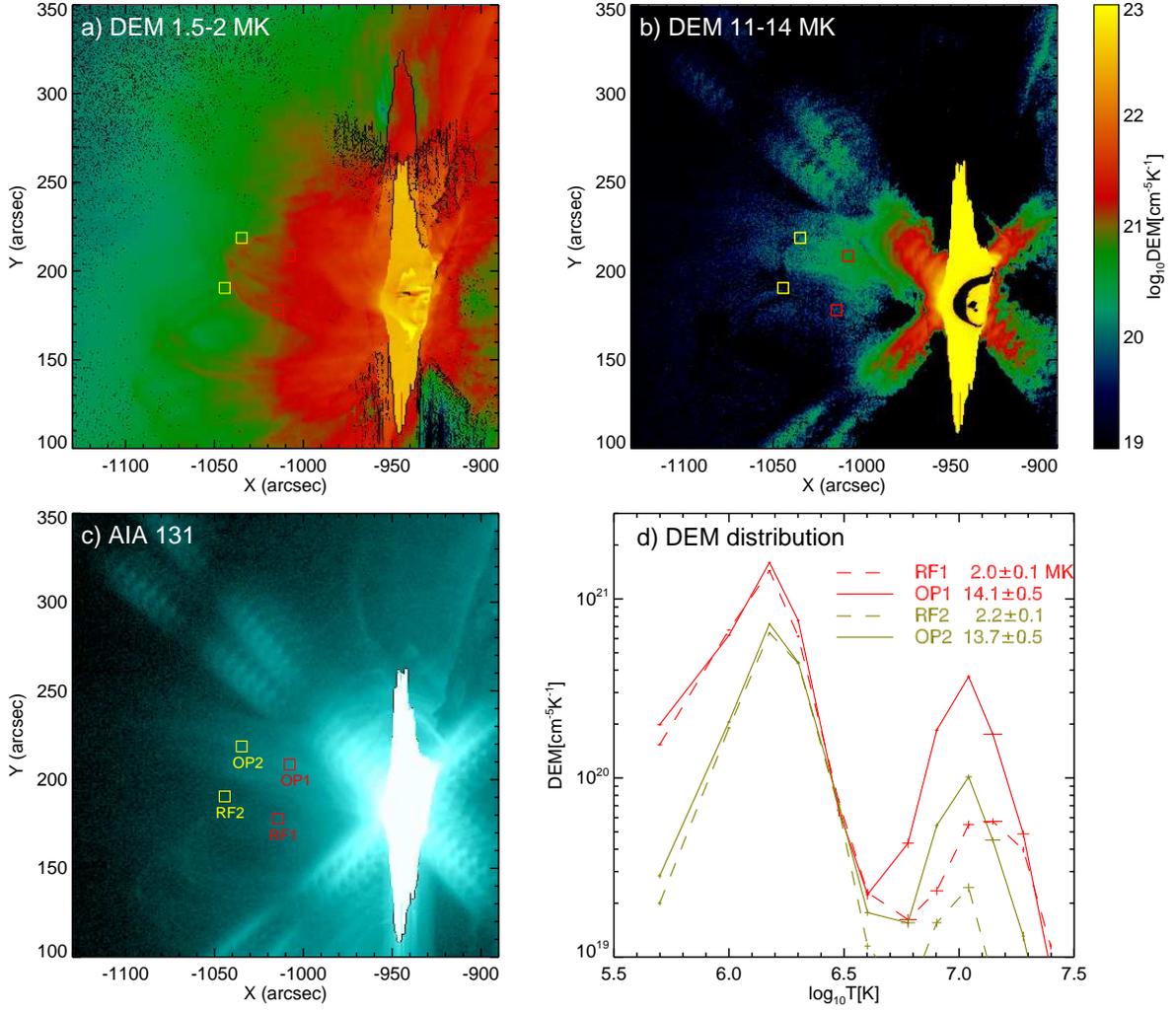}
  \caption{\small DEM results at 16:00~UT. (a,b) DEM maps at temperature bins of 1.5--2 and 11--14~MK. (c) \sdo/AIA 131~\r{A} map at 16:00~UT. (d) Mean DEM results of selected regions as indicated in Panels (a)--(c). The corresponding mean temperatures are given in the unit of MK. 
  \label{fig:dem}}
\end{figure}

In order to confirm the high temperature of OPs, we perform a differential emission measure (DEM) analysis utilizing six AIA EUV passbands data (the optically thick 304~\r{A} is not included), which are further processed to level 1.6 by applying the procedures {\tt aia\_deconvolve\_richardsonlucy} and {\tt aia\_prep} in the Solar Software (SSW). The code developed by \citet{Hannah2012} is used and some results at 16:00~UT are shown in Figure \ref{fig:dem}. The background coronal plasmas generally show up in the DEM maps of temperature below 4~MK, while OPs only in high temperatures exceeding 8~MK (e.g., DEM maps in Figure \ref{fig:dem}(a,b)). In addition, four sub-regions, each including $10\times10$~pixels$^2$, are chosen for a detailed study, with OP1 and OP2 on the plasma jets, RF1 and RF2 nearby at similar altitudes as a reference. The DEM curves exhibit a common double peak distribution \citep[e.g.,][]{Gou2015}. Note that the cool components from OP1 and RF1 are almost the same but the hot one in OP1 is much larger than that in RF1. The DEM curves of OP2 and RF2 have a similar relation to OP1 and RF1. Thus we can conclude that the hot plasmas make a significant contribution to the observed OPs, while in the nearby reference region, the background coronal plasma dominates. We calculate the corrected mean temperature \citep[see details in][]{Gou2015}
\begin{equation}
<T>_h = \frac{\sum\nolimits_{T>4MK} DEM(T) \times T\Delta T}{\sum\nolimits_{T>4MK} DEM(T) \Delta T} ,
\end{equation}
which is only weighted by the hot DEMs above 4~MK, and obtain $\sim$14~MK for OP1 and a slightly lower value for OP2, as shown in Figure \ref{fig:dem}(d). The mean temperatures for RF1 and RF2 in Figure \ref{fig:dem}(d) are only weighted by the cool DEMs (T$<$4~MK), because the small hot component probably comes from scattering. The uncertainties are calculated following the error propagation theory.

The sudden inward motion and the disappearance of the loop leg, as well as the simultaneously upward-moving hot plasmas in the perpendicular direction, provide strong evidence for magnetic reconnection that is driven by the inflow as exhibited by the fast inward swing of the loop leg. The hot OPs, which show both temporal and spatial consistency with the cool inflow, are therefore related to the reconnection outflow roughly moving at the Alfv\'{e}n speed. So we estimate the reconnection rate at that time, $ M_A=v_{in}/v_{A}\approx v_{in}/v_{out} $, as $\sim$0.18 if using the lower inflow speed of $\sim$130~km/s, since it is measured near the reconnection site of the loop, i.e., the region above the flaring loops close to S2, where the hot outflow is observed to move upward. The speeds of the loop motion in the upper regions can be much higher but they are most probably a response to the inflow to the reconnection region below. This interpretation is confirmed by the delay of the start of the inflow swing in the higher region up to $\sim$40~seconds (e.g., compare those in Figure \ref{fig:tempor} (e) and (f)).

\subsection{X-ray Observation}\label{ssec:xray}

\subsubsection{X-ray Fluxes}

Figure \ref{fig:tempor}(a) shows \goes\ and \rhessi\ X-ray fluxes during the impulsive phase of the flare. The \rhessi\ fluxes in the 6--12 and 12--25~keV bands have similar temporal evolution to that from \goes\ 1--8~\r{A}, with gradual time variation and two distinct energy-release episodes. While the fluxes of high energy bands vary rapidly and closely resemble the \goes\ time derivative, consistent with the well-known Neupert effect \citep{Neupert1968,Veronig2002}. Note that right after 16:00~UT, i.e., the time of magnetic reconnection of the loop observed in the AIA EUV channels, \rhessi\ fluxes show strong bursts of high-energy emissions from flare-accelerated particles up to the $\gamma$--ray range, i.e.,~$\geq$~500~keV. This huge energy release and accelerated particles to such high energies further confirms the strong reconnection process.

\subsubsection{RHESSI Imaging}

\begin{figure}[ht]
  \plotone{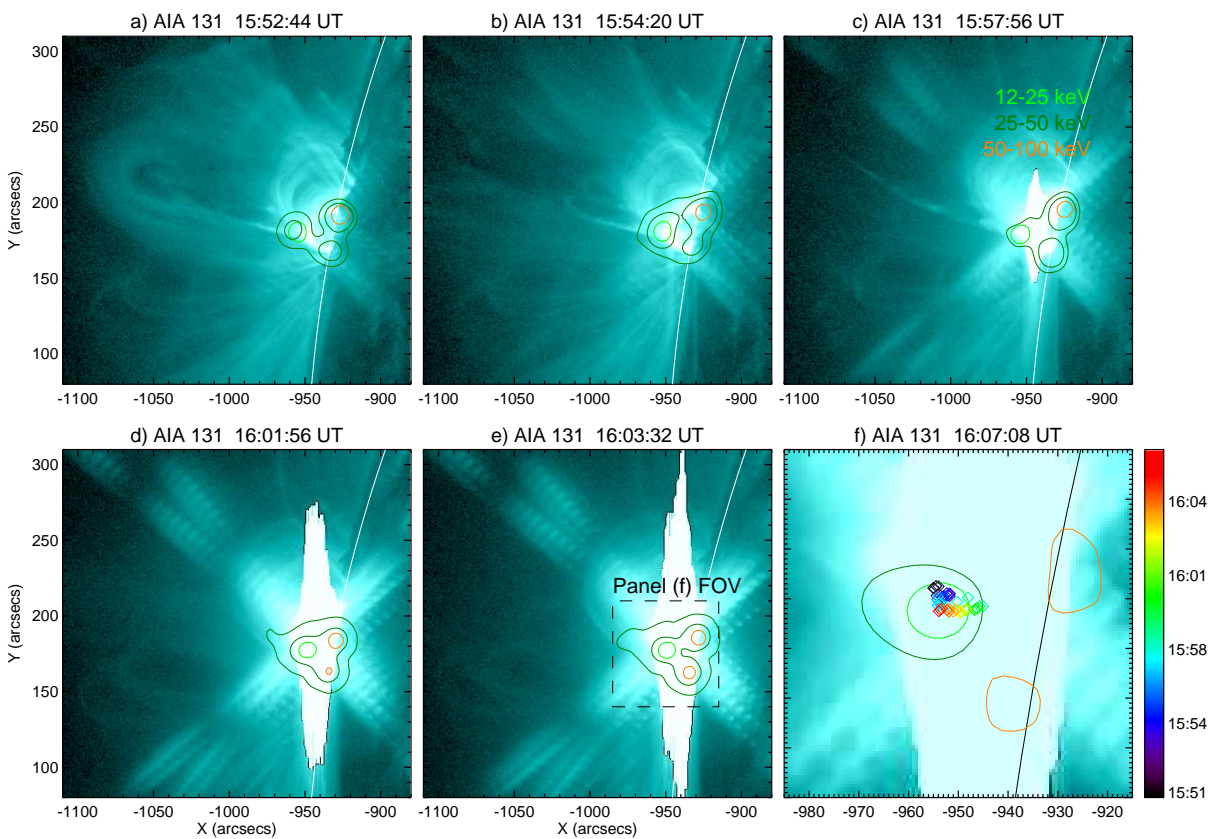}
  \caption{\small \rhessi\ source morphology and evolution, overlaid on the \sdo/AIA 131 \r{A} images. Light green and orange contours are in 12--25 and 50--100 keV bands at 80\% of the maximum in each image. Dark green contours are in 25-50 keV bands at 30\%, 50\% of the peak emission in Panels (a)--(c) before 16:00~UT, at 10\%, 50\% in Panels (d,e), only at 50\% in Panel (f). The locations of 12--25 keV LT sources are denoted by diamond signs in Panel (f) and the temporal evolution from 15:51~UT to 16:07~UT is color coded. The rectangle in Panel (e) indicates the field of view (FOV) of Panel (f). An animation of this figure is available.
  \label{fig:img}}
\end{figure}

We reconstruct \rhessi\ X-ray images with the CLEAN algorithm using front detector segments 3--9, integrating over 20s time intervals. Figure \ref{fig:img} and its animation show the 12--25, 25--50 and 50--100~keV sources which are plotted on corresponding AIA 131~\r{A} images. The 12--25~keV and 50--100~keV emissions mainly come from the flare LT and footpoints respectively, and 25--50 keV emission comes from both of them. One can see an extended source in 25--50~keV above the LT after $\sim$16:00~UT, similar to that in Figure 4 of \citet{MartinezO2014}. 

We determine the centroid of emission above 70\% of the peak flux to study the evolution of the LT source location \citep{Veronig2006}. In Figure \ref{fig:img}(f) we plot the motion locus of \rhessi\ 12--25~keV LT sources. It shows both downward and upward motions with respect to the solar surface. The axis of motion, which is determined by a linear fit to the centroid data, is offset from the solar radial direction by 6$\degree$ toward north. We therefore use the distance above the solar surface as the height of sources, and the evolutions of 6--12 and 12--25~keV sources are shown in Figure \ref{fig:tempor}(c). The 12--25~keV sources are located slightly above the 6--12~keV ones, indicating hotter flare loops are higher, which is consistent with the standard flare model. Note that there is a sharp decrease of the LT source height at around 15:59~UT, in which the 12--25~keV LT drops from $\sim$16~Mm to $\sim$10~Mm within less than 80~seconds with a speed of $\sim$80~km/s. This is similar to the LT altitude decrease observed in the early stage of flares but much faster and stronger, and here it coincides well with the inward swing of the loop leg in AIA 304 \r{A} (see Figure \ref{fig:tempor} (c,e,f)). Thereafter, the LT source continues to move upward, indicating continuous rising of the current sheet and reconnection site.

\subsubsection{X-ray Spectra}
 
We derive the \rhessi\ spectra accumulated over 20s intervals from 15:50~UT to 16:05~UT using front detector 1 and fit them with an isothermal component ({\tt f\_vth}) plus a thick-target bremsstrahlung function ({\tt f\_thick2}) of a power-law electron distribution \citep{Holman2003}. Functions {\tt f\_pileup\_mod} and {\tt f\_drm\_mod} are also used to correct pileup and adjust the detector response model based on the counts around the iron line complex at 6.7 keV. For intervals where the attenuator is in the A3 state (both thick and thin attenuators inserted) additional lines at 8.45 and $\sim$10 keV are also included to compensate for instrumental effects \citep{Phillips2006}. The low energy boundary of fitting is set to 6 keV as there are attenuators in for all intervals, and the high boundaries are set automatically with the keyword {\tt spex\_fit\_auto\_erange }.

\begin{figure}[ht]
  \plotone{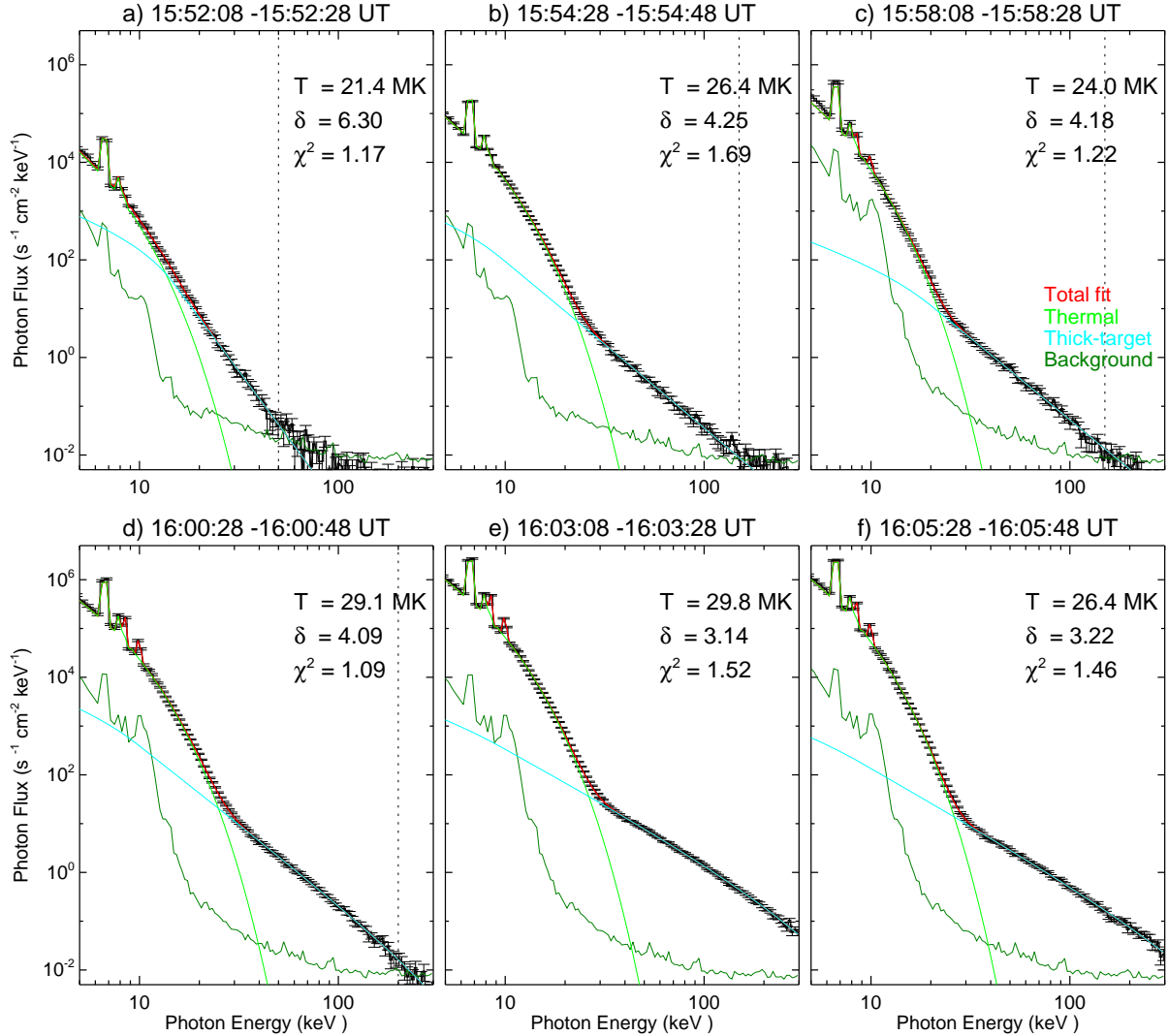}
  \caption{\small \rhessi\ background-subtracted photon spectra (black curves with error bars) and background emission (dark green) for six time intervals. Also shown are the spectral fits, with light green curves for the thermal components, cyan for non-thermal thick-target bremsstrahlung, and red for the sum. Temperature (in the unit of MK) of thermal plasmas, electron distribution index $\delta$ from thick-target bremsstrahlung model, and reduced $\chi^2$ of the fits are denoted in each panel. The vertical dotted lines show the high energy boundaries of fitting.
  \label{fig:spec}}
\end{figure}

Figure \ref{fig:spec} shows some examples of background subtracted \rhessi\ spectra and fitting results. The obtained spectral fits are acceptable with reduced $ \chi ^2\sim$1. After 16:00~UT the thermal temperatures increase significantly and non-thermal particles come from much higher energies. The detailed temporal evolution of fitting parameters, i.e., emission measure (EM) and temperature (T) from the thermal component, and the power-law index ($\delta$) of the electron distribution derived from the thick-target bremsstrahlung model, is shown in Figure \ref{fig:tempor}(b). One can see that T (red cross symbols) exhibits two distinct peaks and the second one, which occurs after 16:00~UT, is higher with T $\approx$ 30 MK, indicating a superhot component \citep{Lin1981} during the second episode of energy release. A thermal fit to \goes\ data (black and red lines) reveals a similar trend both in EM and T, although the obtained EM is larger and T is smaller due to the preferential response to relatively cooler plasma compared to \rhessi. The spectral index $\delta$ of the flare-accelerated electrons generally changes between 5 and 4 before 16:00~UT, while it decreases to $\sim$3 during the second energy-release episode, indicating a soft-hard-harder (SHH) evolution, which could result from magnetic trapping of energetic particles in the corona \citep{Metcalf1999}.

\section{Summary and Discussion} \label{sec:sum}

We study the X2.8 flare on 2013 May 13 which exhibits unusual two-step magnetic reconnection as well as two distinct episodes of energy release. The first episode is associated with the erupting flux rope, similar to most flare-CME events that show a close correlation between the impulsive phase of CME acceleration and the flare energy release  \citep[e.g.,][]{Zhang2001,Temmer2008}. However, the interesting and anomalous behavior in this event is the second-step magnetic reconnection of one large-scale loop in the aftermath of the eruption, which is accompanied by another, even stronger energy-release episode. We present clear observations of a plethora of magnetic reconnection signatures associated with the second phase. The magnetic inflow of the loop leg is suddenly swept into the reconnection region at a very high speed of $\sim$130 km/s, faster than previous observations which generally range from 5 to 100 km/s \citep[e.g.,][]{Yokoyama2001,Lin2005,Takasao2012,Su2013}. Together with the corresponding hot outflow moving upward from that region approximately at Alfv\'{e}n speed, we obtain a reconnection rate of $\sim$0.18 at that time. \rhessi\ observations show significant energy release and hard spectra ($\delta\approx$ 3) with emissions up to the $\gamma$--ray range, as well as a second larger temperature peak up to T$\approx$30 MK, indicative of a superhot component. 

The contraction and oscillation of the coronal loops in AIA 171 \r{A} images can be interpreted as the coronal implosion, according to which the force balance of the loop is disrupted due to a reduction in the magnetic pressure resulted by magnetic energy release associated with the eruption \citep{Hudson2000}. This consequent decrease of magnetic pressure around the reconnection site can also explain the inward swing of the loop leg inflow in AIA 304 \r{A} and simultaneously the sharp altitude decrease of the \rhessi\ LT source. Furthermore, the loop leg that is quickly sucked into the reconnection site causes a sudden increase in the inflow speed and magnetic flux, which can inevitably overshoot and lead to magnetic flux pileup around the current sheet. As a result, the current sheet is expected to become thinner and longer, causing a possible extension in both upward and downward directions, which is also consistent with the altitude decrease of the LT source. In return, the thinning and extension of the current sheet may trigger the tearing mode instability \citep{Furth1963}, which would lead to multiple O-type and X-type reconnection sites, therefore enhancing the reconnection rate (e.g., \citealt{Lin2015} and references therein). 

\acknowledgments
We thank the \sdo\ and \rhessi\ teams for the open data policy and the development of data analysis software. T.G. acknowledges the support from the China Scholarship Council (CSC) under file No.201606340097. A.M.V. and A.H.-P. gratefully acknowledge the Austrian Science Fund (FWF): P27292-N20. R.L. acknowledges the support by NSFC 41474151 and the Thousand Young Talents Program of China.

\clearpage

\end{document}